\documentclass[12pt,preprint]{aastex}
\def\myputfigure#1#2#3#4#5%
{\hskip0.03\textwidth\vskip#5pt%\hfill
\makebox[0pt]{\hskip#2in
\includegraphics[width=#3\textwidth]{#1}}\vskip#4pt\hfill}


\def\fun#1#2{\lower3.6pt\vbox{\baselineskip0pt\lineskip.9pt
\ialign{$\mathsurround=0pt#1\hfil##\hfil$\crcr#2\crcr\sim\crcr}}}
\def\lap{\mathrel{\mathpalette\fun <}}
\def\gap{\mathrel{\mathpalette\fun >}}

\def\mass{{\cal M}}
\def\en{{\cal M}}

\def\msun{{\mass_\odot}}

\def\beq{\begin{equation}}
\def\eeq{\end{equation}}

\def\mh{M_{\bullet}}

\lefthead{Chaotic Loss Cones}
\righthead{}

\begin{document}

\title{Chaotic Loss Cones, Black Hole Fueling and the $\mh-\sigma$ Relation}

\author{David Merritt$^1$ and M. Y. Poon$^2$}

\affil{$^1$Department of Physics and Astronomy, Rutgers University,
New Brunswick, NJ 08903; \\
$^2$ Harvard/Smithsonian Center for Astrophysics, 60 Garden Street,
Cambridge, MA 02138; \\
merritt@physics.rutgers.edu, mpoon@cfa.harvard.edu}

\begin{abstract}
In classical loss cone theory, stars are supplied to a
central black hole via gravitational scattering onto 
low angular momentum orbits.
Higher feeding rates are possible if the gravitational
potential near the black hole is non-axisymmetric and the
orbits are chaotic.
Motivated by recently published, self-consistent models,
we evaluate rates of stellar capture and disruption in triaxial nuclei.
Rates are found to substantially exceed those in collisionally-resupplied
loss cones, as long as an appreciable fraction of the orbits are centrophilic.
The mass captured by a black hole after a given time in a steep
($\rho\sim r^{-2}$) nucleus
scales as $\sigma^5$
with $\sigma$ the stellar velocity dispersion,
and the accumulated mass in $10^{10}$ yr is of the correct order to
reproduce the $\mh-\sigma$ relation.
Triaxiality can solve the ``final parsec problem'' of decaying
black hole binaries by increasing the flux of stars into the
binary's loss cone.
\end{abstract}

\section{Introduction}

Fueling of active galactic nuclei (AGNs) or quasars requires
accretion at rates of $\dot{M}\approx 1.5 \epsilon_{0.1}^{-1}L_{46}
\msun \mathrm{yr}^{-1}$ onto the central supermassive black hole (SBH), 
where $L_{46}$ is the energy output in units of $10^{46}$ ergs s$^{-1}$ 
and $\epsilon_{0.1}$ is the mass conversion efficiency in units of
its canonical value $0.1$.
The fueling problem is usually broken into three parts:
what is the fuel; how is it channeled into the SBH;
and how does it radiate a substantial fraction of
its energy before disappearing down the hole?
This paper addresses the first two questions.
Galactic spheroids have ample supplies of both stars and gas,
but it is difficult to come up with mechanisms that can
extract almost all of a mass element's orbital angular momentum
in a few crossing times,
as required if the matter is to find its way into the event
horizon of the black hole.
Gas release through tidal disruption of stars is a possible 
fueling mechanism \citep{Hills75,FR76},
but the rate at which stars are scattered onto low angular
momentum orbits by gravitational encounters is too low
to reproduce observed luminosities
(Young, Shields \& Wheeler 1977; Frank 1978).
The fact that most AGNs are in spiral galaxies suggests that
interstellar gas provides the bulk of the fuel.
Gas can be driven into the SBH by
torques from non-axisymmetric potential perturbations, due 
to stellar bars or to transient distortions
of the potential during mergers or accretion events
(Shlosman, Begelman \& Frank 1990).

As a number of authors have pointed out, {\it stellar} feeding rates
are also enhanced by the presence of non-axisymmetric perturbations
\citep{NS83,GB85}.
A barlike or triaxial potential is populated mainly by box
orbits, and a star on a box orbit passes near the center once 
per crossing time.
If a SBH is added, most of these ``centrophilic'' orbits 
become chaotic due to large-angle deflections by the SBH
\citep{VM98}.
Estimates of feeding rates due to centrophilic orbits
in triaxial potentials are several times larger than
rates due to scattering onto eccentric
orbits in the spherical or axisymmetric geometries
\citep{GB85}.
While such an enhancement is significant, it is still not
enough to explain the high luminosities of AGNs and quasars.

A number of factors motivated us to re-open the question of 
stellar fueling rates in non-axisymmetric nuclei.
The $\mh-\sigma$ relation demonstrates a tight link between
SBH masses and the kinematics of their stellar spheroids.
While the origin of the relation is still uncertain,
some scenarios postulate a role for stellar feeding
(e. g. Zhao, Haehnelt \& Rees 2002).
Capture of stellar-mass objects by SBHs may provide an
important source of signals for gravitational wave detectors
like LISA \citep{Hughes01}.
Imaging of the centers of galaxies on scales of $\sim 10$ pc 
reveals a wealth of features in the stellar
distribution that are not consistent with axisymmetry,
including bars, bars-within-bars, nuclear spirals,
and other misaligned structures
\citep{Wozniak95,Rest01,Peng02,ES02}.
The steep power-law dependence of stellar density on
radius near the centers of many galaxies revealed by HST
\citep{Crane93,Gebhardt96}
implies higher stellar feeding rates than in older models 
\citep{NS83,GB85} which postulated constant-density cores.
Triaxiality has recently been shown
to be sustainable in numerical models for black-hole nuclei
(Poon \& Merritt 2001, 2002, 2003, hereafter Papers I-III).
Furthermore such models can contain a large population
of centrophilic -- typically chaotic -- orbits,
as high as $\sim 75\%$.

This paper, the fourth in a series on the dynamics of
triaxial black-hole nuclei, examines the behavior
of centrophilic orbits and the implications for black hole
feeding.
Test-particle integrations in fixed potentials are combined
with knowledge of the orbital population
in the self-consistent models to infer the rate at which
stars would be supplied to the central SBH.
We confirm and extend the results of earlier authors,
who investigated stellar feeding rates in triaxial nuclei with 
constant-density cores.
We find that stellar capture rates in steep power-law nuclei
can be orders of magnitude greater than in nuclei
with cores; in fact, feeding rates can meet or even exceed 
the so-called ``full loss cone'' rate, i.e. 
the capture rate in a spherical or axisymmetric
galaxy in which the loss cone is continuously repopulated.
The accretion rate in a dense, $\rho\sim r^{-2}$ nucleus
scales as $\sigma^5$ and the mass accumulated
in $10^{10}$ yr
is of the right order to reproduce 
the $\mh-\sigma$ relation.

A number of consequences follow from such high rates of stellar feeding.
Tidal disruption events at the present epoch
could be significantly more frequent than in models
based on collisional loss-cone repopulation.
The decay rate of a binary SBH could be signficantly enhanced compared
with the rate in a spherical or axisymmetric nucleus, allowing 
binary SBHs
to overcome the ``final parsec problem'' \citep{MM03a} and coalesce
by emission of gravitational radiation.
Even if long-lived triaxiality should turn out to be rare, we show that
transient departures from axisymmetry during mergers or galaxy interactions
could induce feeding at rates approaching those inferred in AGNs.

\section{Models and Units}

The models of Papers II and III were based on the density law
\begin{mathletters}
\begin{eqnarray}
\rho_{\star} &=& \rho_0 m^{-\gamma},\\
m^2 &=& \frac{x^2}{a^2}+\frac{y^2}{b^2}+\frac{z^2}{c^2}
\label{eq:den}
\end{eqnarray}
\end{mathletters}
with $\gamma= (1,2)$ and $c/a=0.5$.
The outer boundary (an equipotential surface) was chosen to 
contain roughly $100 (20)$ times
the black hole mass for $\gamma=1 (2)$.
Non-evolving solutions for both values of $\gamma$ were found for 
$T=0.25$ (nearly oblate) and $T=0.50$ (maximal triaxiality); 
$T=(a^2-b^2)/(a^2-c^2)$ is the triaxiality index.
Models with $T=0.75$ (nearly prolate) evolved rapidly into precisely 
axisymmetric shapes and are not considered here.
The black hole was represented by a central point with 
unit mass; the long-axis scale length $a$ and the constant of gravitation $G$ 
were also set to unity.
The scale-free nature of the mass distribution allowed
us to set $\rho_0=1$ without loss of generality.

Our focus is on models which include chaotic orbits.
Four such models were presented in Paper III.
Table 1 gives the axis ratios and the mass fractions on chaotic,
tube and pyramid (box) orbits for these four models.
(For definitions of the orbit families, see Papers I-III.)
Chaotic orbits and tube orbits contributed roughly equally to the
total mass in the self-consistent solutions, 
with regular box orbits a distant third.
We therefore do not treat the box orbits separately from the 
chaotic orbits in what follows.
The mass fraction in ``centrophilic'' orbits (chaotic orbits and pyramids)
was $\sim 50\%$ in all models.

It is convenient to define equivalent spherical models for the
stellar mass distribution; as shown in Paper III, these spherical models
have energy distributions $N(E)$ similar to those of the triaxial
models and are useful when scaling the results found here to
real galaxies.
The equivalent spherical models have $\rho_*(r) = (r/\delta)^{-\gamma}$
with $\delta=(abc)^{1/3}$, or $\delta=0.734 (0.767), T= 0.5 (0.25)$.
Their gravitational potentials in model units are
\begin{mathletters}
\begin{eqnarray}
\Phi_*(r) &=& 2\pi \delta r,\ \ \ \ \gamma=1 \\
&=& 4\pi \delta^2\left[\ln\left({r\over \delta}\right) - 1\right],\ \ \ \ \gamma=2.
\end{eqnarray}
\end{mathletters}
The additive constants in these expressions for the potential have been chosen
in the same way as for the triaxial models (see Paper I); we note that,
for $\gamma=2$, the zero point of the potential occurs
at a radius containing $4\pi e \delta^3$ times the black hole mass $\mh$,
or $13.51 (15.41) \mh$ for $T=0.5 (0.25)$.
Note that the black hole has been omitted in these expressions for the
potential.

We define $r_h$ to be the radius in the spherical model containing
a mass in stars equal to twice the black hole mass (cf. Merritt 2003).
This definition is equivalent to the standard one, $r_h=G\mh/\sigma^2$, 
when $\gamma=2$, the singular isothermal sphere.
For $\gamma=1$, $r_h=(\pi \delta)^{-1/2}=0.659 (0.643)$ for $T=0.5(0.25)$, 
while for $\gamma=2$, $r_h=(2\pi \delta^2)^{-1}=0.296 (0.270)$.
The 1D stellar velocity dispersion $\sigma$ is 
equal to $\sqrt{2\pi}\delta=1.84 (1.92)$ in model units for $\gamma=2$.
The velocity dispersion in a nucleus with $\gamma=1$ depends
on radius and is zero at the center in the absence of a black hole
\citep{Dehnen93}; hence we do not quote a value for $\sigma$ when 
$\gamma=1$.

We define $E_h\equiv\Phi(r_h)$.
In model units, $E_h=3.04 (3.10)$ for $\gamma=1$ and 
$-12.93 (-15.01)$ for $\gamma=2$.

The results presented below can be related to real galaxies
using the following scale factors for mass, length and time,
derived from the spherical models just defined.
Numerical constants, when given, are for 
$T=0.5$ and differ only slightly for $T=0.25$.
\begin{equation}
\left[M\right] = \mh 
\end{equation}
\begin{mathletters}
\begin{eqnarray}
\left[L\right] &=& \left(\pi\delta\right)^{1/2}r_h\approx 152\ {\rm pc} \left({r_h\over 100\ {\rm pc}}\right),\ \ \ \ \gamma=1 \\
&=& \left(2\pi\delta^2\right)r_h\approx
37.4\ {\rm pc} \left({\sigma\over 200\ {\rm km\ s}^{-1}}\right)^{-2}\left({\mh\over 10^8\msun}\right),\ \ \ \  \gamma=2 
\end{eqnarray}
\end{mathletters}
\begin{mathletters}
\begin{eqnarray}
\left[T\right] &=& \left(\pi\delta\right)^{3/4}\sqrt{r_h^3\over G\mh}
\approx 2.79\times 10^6 {\rm yr} \left({r_h\over 100\ {\rm pc}}\right)^{3/2} \left({\mh\over 10^8\msun}\right)^{-1/2},\ \ \ \  \gamma=1 \\
&=& \left(2\pi\delta^2\right)^{3/2}\sqrt{r_h^3\over G\mh} 
= 3.30\times 10^5 {\rm yr} \left({\sigma\over 200\ {\rm km s}^{-1}}\right)^{-3} \left({\mh\over 10^8\msun}\right),\ \ \ \ \gamma=2.
\label{eq:defT2}
\end{eqnarray}
\end{mathletters}
In equation (\ref{eq:defT2}), the relation $\sigma^2=G\mh/r_h$ was used.

The tidal disruption radius $r_t$ of a $10^8\msun$ black hole is 
roughly equal to its Schwarzschild radius $r_s=2G\mh/c^2$.
In model units, the Schwarzschild radius is
\begin{mathletters}
\begin{eqnarray}
r_s &=& \sqrt{1\over\pi\delta} {2G\mh\over c^2 r_h}\approx6.31\times 10^{-8}\left({r_h\over 100\ {\rm pc}}\right)^{-1}\left({\mh\over 10^8\msun}\right), \ \ \ \ \gamma=1 \\
&=& {1\over \pi\delta^2}{G\mh\over c^2r_h}\approx0.591\left({\sigma\over c}\right)^2\approx 2.63\times 10^{-7}\left({\sigma\over 200\ \mathrm{km\ s}^{-1}}\right)^2,\ \ \ \ \gamma=2.
\label{eq:sch}
\end{eqnarray}
\end{mathletters}

In order to keep the notation as simple as possible, 
the same symbols will be used for both dimensional and
dimensionless quantities in what follows.
When not noted explicitly,
the distinction will be clear from the form of the 
equations.

\section{Pericenter Distributions}

Chaotic orbits pass near the central singularity
once per crossing time, although the pericenter
distance -- the distance of closest approach to the black
hole -- varies quasi-randomly from passage to passage.
We computed the statistics of pericenter passages by carrying
out integrations of chaotic orbits for $10^5 T_D$ in 
each of the four triaxial potentials;
$T_D$ is the energy-dependent dynamical time defined in Paper I.
To speed the integrations, the potential and forces due
to the stars (equation \ref{eq:den}) were expressed in terms
of a truncated basis set \citep{HO92};
Ten radial and angular functions were used.
The routine RADAU \citep{Hairer96} was used for the integrations.
Pericenter distances were computed by locating two times
$(t_1, t_2)$ between which the distance to the black hole reached a minimum
and interpolating the solution on a fine grid in $t_1<t<t_2$
to find the distance of closest approach.

Figure 1 shows the cumulative distribution of pericenter distances
for two chaotic orbits integrated
in the triaxial potentials with $\gamma=(1,2)$ and $T=0.5$.
The number of central encounters per unit time with pericenter 
distances less than $d$, $N(r_p<d)$, is nearly linear with $d$ over 
its entire range.
The linear dependence was observed to extend down to pericenter distances 
of $10^{-6}$ or smaller, of order the tidal disruption radius
in model units.
The approximately linear dependence of $N_E$ on $d$,
combined with the gravitational focussing equation
\beq
r_b^2=d^2\left(1+{2G\mh\over V^2d}\right) \approx {2G\mh\over V^2}d,
\eeq
with $r_b$ the impact parameter,
implies $N(r_p<r_b)\propto r_b^2$, 
i. e. a nearly uniform filling of the two-dimensional cross section that
defines the ``throat'' of the chaotic orbit (cf. Gerhard \& Binney 1985).

Chaotic orbits were observed to quickly fill
the configuration-space region accessible to them, and one therefore
expects that the statistical properties of one chaotic orbit
at a given energy will be the same as those of any other chaotic
orbit at the same energy.
We verified for a number of chaotic orbits that $N(r_p<d)$
is indeed only weakly dependent on the starting point
for fixed energy $E$.
Hence we can write $N(r_p<d)=N_E(r_p<d)$, the rate of pericenter passages
for a star on a chaotic orbit of energy $E$.

We computed $N_E(r_p<d)$ at the energies associated with each of
the mass shells defined in Paper II.
Let $A(E)d$ be the rate at which a single star on a chaotic orbit
of energy $E$ experiences pericenter passages with $r_p<d$.
The dependence of $A$ on $E$ in the four triaxial potentials
is shown in Figure 2.
For $\gamma=2$, an exponential fits the data well:
\beq
\ln A \approx a + bE,
\eeq
with
\begin{mathletters}
\begin{eqnarray}
&& a = -0.603,\ \  b=-0.290 \ \  (T=0.5) \\
&& a = -0.655,\ \  b=-0.275 \ \ (T=0.25).
\end{eqnarray}
\end{mathletters}
We note that $b\approx -1/2\pi\delta^2$.
Henceforth we set $b\equiv -1/2\pi\delta^2$;
thus $A\propto e^{-E/\sigma^2}$ and $A$ scales with radius as 
$\sim r^{-2}$.
This simple scaling will be useful in what follows.
In physical units,
\beq
A(E) \approx 1.2 {\sigma^5\over G^2\mh^2} e^{-(E-E_h)/\sigma^2}.
\label{eq:aofe}
\eeq
Figure 2 shows that this relation holds even at energies
$E\lap E_h$.

For $\gamma=1$, a power law provides a good fit in the energy
range $E\gap E_h$:
\beq
\ln A \approx a + b\ln E,
\eeq
with
\begin{mathletters}
\begin{eqnarray}
&&a=1.734,\ \ b=-1.388 \ \ (T=0.5) \\
&&a=1.869,\ \ b=-1.357 \ \ (T=0.25). 
\end{eqnarray}
\end{mathletters}
In physical units,
\beq
A(E) \approx 0.7 \sqrt{G\mh\over r_h^5} \left(E\over E_h\right)^{-1.4}.
\eeq
Thus $A\sim r^{-1.4}$, $r\gap r_h$.

The pericenter distribution at any $E$ is also
characterized by a second quantity,
the {\it maximum} pericenter distance $r_{p,max}(E)$
reached by chaotic orbits of energy $E$.
This distance corresponds roughly to the width of the ``throat''
of a box orbit in an integrable triaxial potential.
We find, for $\gamma=2$,
\beq
\ln r_{p,max} \approx c + dE,
\eeq
with
\begin{mathletters}
\begin{eqnarray}
&&c = -0.43,\ \  d = 0.146\ \ (T=0.5)\\
&&c = -0.36,\ \  d = 0.141\ \ (T=0.25).
\end{eqnarray}
\end{mathletters}
We note that $d=1/4\pi\delta^2$ to within its uncertainties; thus
\beq
r_{p,max} \approx 0.3 r_h e^{(E-E_h)/2\sigma^2}
\eeq
and $r_{p,max}$ scales approximately linearly with apocenter
distance, $r_{p,max}\approx 0.2 r_{apo}$.

For $\gamma=1$, we find, for $E\gap E_h$,
\beq
\ln r_{p,max} \approx c + d\ln E,
\eeq
with
\begin{mathletters}
\begin{eqnarray}
&&c = -1.86,\ \  d = 0.86\ \ (T=0.5)\\
&&c = -2.03,\ \  d = 0.90\ \ (T=0.25).
\end{eqnarray}
\end{mathletters}
This implies
\beq
r_{p,max} \approx 0.60 r_h \left({E\over E_h}\right)^{0.88}
\eeq
and again $r_{p,max}$ scales roughly linearly with $r_{apo}$.

\section{Feeding Rates}

The feeding rates implied by Figure 2 are high compared with
those in diffusive loss cone models;
in fact they are comparable to the maximum possible capture rates
in spherical models, the so-called ``full loss cone'' rate.
We illustrate this by considering the rate at which stars on 
a single orbit pass within a distance $r_t$ of the center.
In the triaxial models, this rate is $\sim A(E)r_t$.
Setting $E$ to the energy of a circular orbit at $r_h$,
taking $r_t\approx 2\times 10^{-7}$ in model units 
(equation \ref{eq:sch} for $\gamma=2$),
and using the expressions given above for $A(E)$,
we find a rate of $\sim 2\times 10^{-6} T_D^{-1}$.
In other words, roughly $10^6$ orbital
periods are required for a star on the edge of the black hole's
sphere of influence to pass within its tidal radius.
In a spherical model with a full loss cone, 
a fraction $\sim r_tr_h/r^2$ of the stars at $r$
come within $r_t$ each orbital period;
hence the mean capture rate from stars near $r_h$ is
$\sim (r_t/r_h) T_D^{-1}$ or $\sim 1.5\times 10^{-6} T_D^{-1}$.
In the spherical geometry, the net rate of pericenter passages
is due to a small population of stars (those with sufficiently
low angular momenta) being lost over
a short period of time (an orbital period).
In the triaxial geometry,
a large population of stars (all stars on chaotic orbits)
are supplied to the center over a longer period
of time ($\sim 10^5$ periods).
Of course, in the absence of loss cone repopulation,
capture rates in the spherical geometry would drop to zero
after a single orbital period.

Next we compute the full feeding rates in the triaxial models.
Let $\en_c(E,t)dE$ be the mass in stars on chaotic orbits 
with energies from $E$ to $E+dE$; the time dependence reflects a
possible loss of stars due to tidal disruption or capture.
If the capture or disruption radius is $r_t$,
the energy-dependent loss rate is
\begin{mathletters}
\begin{eqnarray}
\dot \en_c(E,t)dE &=& -r_tA(E)\en_c(E,t) dE \\
&=& -r_tA(E)\en_c(E,0)e^{-A(E)r_t t} dE
\label{eq:capture}
\end{eqnarray}
\end{mathletters}
and the total capture rate from chaotic orbits at all energies is
\beq
\dot M(t) = r_t\int A(E)\en_c(E,0)e^{-A(E)r_t t} dE.
\eeq

We evaluated these expressions in two ways.
$\en_c(E,0)$ can be computed directly from the orbital weights
in the Schwarzschild solutions (cf. Figure 3 of Paper III).
Alternatively,
smooth approximations to $\en_c(E,0)$ can be
computed using the equivalent spherical models defined above.
In the latter case, we define $\en_c(E,0)\equiv f_c(E)\en(E)$ with 
$\en(E)$ the energy distribution in the equivalent spherical model and
$f_c(E)$ the fraction of stars at energy $E$ assumed to be
on chaotic orbits.
For $\gamma=1$, the energy distribution in the spherical geometry is
\begin{mathletters}
\beq
\en(E) = {8\over 35} {r_h^2\over G^2\mh} E
\eeq
or, in model units,
\beq
\en(E) = {8\over 35\pi\delta}E.
\label{eq:nofe1}
\eeq
\end{mathletters}
For $\gamma=2$, we have
\begin{mathletters}
\beq
\en(E) = {2\sqrt{6}\over 9}{r_h\over G} e^{(E-E_h)/2\sigma^2}
\eeq
or, in model units,
\begin{eqnarray}
\en(E)= {2\sqrt{6}\over 9} e\delta e^{E/4\pi\delta^2}.
\label{eq:nofe2}
\end{eqnarray}
\end{mathletters}
Since these approximations to $\en(E)$ were derived from 
expressions for the potential that exclude the
central point mass, they are only strictly valid outside the black hole's 
sphere of influence, $E\gap E_h$.
However the expression for $\gamma=2$ turns out to be reasonably correct even
at lower energies (cf. Figure 3 of Paper III), 
a result that will be used below.

Figure 3 shows $\dot\en_c(E,0)$ computed in both ways.
The chaotic orbit fraction $f_c(E)$ was
set to a constant, $\overline{f_c}$, in each model; 
no attempt was made to match
the detailed energy dependence of the chaotic orbital populations in the 
Schwarzschild solutions.
Nevertheless the fits are reasonably good, particularly for $\gamma=2$.
For $\gamma=1$, the orbital distributions in the Schwarzschild
solutions are ``noisy'' but the analytic expression does a reasonable
job of reproducing the mean dependence.
The values of $\overline{f_c}$ used
in Figure 3 were $\overline{f_c}=0.25$ ($\gamma=2, T=0.5$ and $0.25$),
$\overline{f_c}=0.5$ ($\gamma=1, T=0.5$) and 
$\overline{f_c}=0.4$ ($\gamma=1, T=0.25$).
These values are slightly smaller than the overall chaotic mass
fractions in the numerical solutions (Table 1);
the reason is that the Schwarzschild solutions put  
a large number of orbits in the outermost shells to compensate
for the hard outer edge, and are correspondingly
depleted at low energies (Figure 3 of Paper III).

Also shown in Figure 3 are the feeding rates predicted by a 
spherical, ``full loss cone'' model.
The full-loss-cone feeding rate is
\begin{mathletters}
\begin{eqnarray}
\dot M_{full}(E) dE &=& {\en_{lc}(E)\over P(E)} dE \\
&=& 4\pi^2 f(E) J_{lc}^2(E) dE
\label{eq:full}
\end{eqnarray}
\end{mathletters}
where $\en_{lc}(E)$ is the number of stars in the spherical
model with pericenters below $r_t$, $f(E)$ is the isotropic
distribution function, $P(E)$ is the radial period, and
$J_{lc}$ is the angular momentum of a star with pericenter $r_t$,
$J_{lc}^2=2r_t^2(E-\Phi(r_t))\approx 2G\mh r_t$.
Equation (\ref{eq:full}) describes the capture rate in a spherical
model assuming that all of the orbits remain fully populated.
Figure 3 shows that the capture rates in the triaxial models
are similar to the full-loss-cone rates in the equivalent spherical
models, and can even exceed them at high energies.

In a real spherical galaxy, the high feeding rates 
corresponding to a full loss cone would
persist for just a single orbital period at each energy.
In the triaxial models, by contrast, the capture rate from
chaotic orbits 
decreases only by a factor $\sim \exp\left[A(E)P(E)r_t\right]$ in one
orbital period.
At energies $\gap E_h$, this factor is negligible meaning that
feeding rates will remain large for many orbital periods.

The evolution of $\dot\en_c(E,t)$ is shown in Figure 4.
Here we have set the capture radius $r_t$ to its smallest possible
value, $r_t=r_s$, using equation (\ref{eq:sch}).
For $\gamma=1$, this requires a choice for $\mh$ and $r_h$;
we took $\mh=10^8\msun$ and $r_h= 100$ pc.
For $\gamma=2$, only $\sigma$ needs to be specified; we took
$\sigma = 200$ km s$^{-1}$.
The analytic approximations, equations (\ref{eq:nofe1}) and (\ref{eq:nofe2}), 
were used for $M_c(E,0)$ and we set $\overline{f_c}=0.5$.
Figure 4 shows that, for $\gamma=1$, the capture rate is low enough that no
significant changes occur in $\en_c(E,t)$ even at times as great
as $t=10^6$, corresponding to $\sim 10^{12}$ yr using the scaling
of equation (\ref{eq:defT2}).
For $\gamma=2$, the capture rate is higher and 
changes in $\en_c(E,t)$ begin to occur at energies above $E_h$ 
for $t\gap 10^5$, corresponding to $\sim 10^{10}$ yr.

The total capture rate from orbits at all energies
is $\dot M(t)=\int\dot\en_c(E,t)dE$.
For $\gamma=1$ and $r_t\approx r_s$, the captured mass is
a small fraction of $\mh$ and $\dot m$ is nearly independent
of time.
Thus we can write
\begin{mathletters}
\begin{eqnarray}
\dot M&\approx& \int_{E_1}^{E_2}\dot\en_c(E,0) dE =
r_t\int_{E_1}^{E_2}A(E)f_c(E)\en(E)dE\\
&\approx& {8\over 35\pi\delta}e^a r_t\int_{E_1}^{E_2}f_c(E) E^{1+b}dE
\end{eqnarray}
\end{mathletters}
in model units.
As lower integration limit we take $E_1=0$; this is reasonable
since the value of the integral
is not strongly dependent on the lower cutoff.
However since $1+b\approx -0.4$, the integral diverges
for large $E_2$.
This corresponds physically to the fact that a $\rho\propto r^{-1}$ 
nucleus can not extend indefinitely.
Nuclear density profiles in real, weak-cusp galaxies exhibit
a break at a radius $r_b\approx 10 r_h$.
We define $E_b\equiv\Phi(r_b)$ and set $E_2=E_b$,
giving, in model units,
\beq
\dot M = {8\over 35(2+b)\pi\delta}e^a r_t\overline{f_c} E_b^{2+b}.
\eeq
Setting $r_t=r_s$, this becomes, in physical units,
\beq
\dot M = \left(1.77,1.97\right)\times 10^{-5}\msun\ {\rm yr}^{-1}
\overline{f_c} 
\left({r_h\over 100\ {\rm pc}}\right)^{-5/2} \left({\mh\over 10^8\msun}\right)^{5/2} \left({r_b\over 10r_h}\right)^{\left(0.61,0.64\right)}
\label{eq:feed1}
\eeq
where the number pairs in parentheses refer to $T=(0.5,0.25)$ respectively.
We note that the feeding rate is almost the same for the two
values of the triaxiality index, suggesting a weak dependence
of $\dot\en$ on the degree of departure from axisymmetry, for
a given chaotic mass fraction.
The feeding rates of equation (\ref{eq:feed1}) imply a captured
mass of $\lap 0.01\mh$ over a Hubble time.

For $\gamma=2$, the feeding rates are much higher, and
the captured mass can be a large fraction of $\mh$
even when $r_t$ is as small as $r_s$.
Hence the population of chaotic orbits is significantly depleted
and we can not ignore the time dependence of $\dot m$.
We have
\begin{mathletters}
\begin{eqnarray}
\dot M(t) &=& r_t\int A(E)\en_c(E,0)e^{-A(E)r_tt}dE\\
&=&{2\sqrt{6} e\delta\over 9} e^a r_t\int f_c(E) e^{-E/4\pi\delta^2} \exp\left({r_tte^{a+bE}}\right) dE
\end{eqnarray}
\end{mathletters}
in model units.
It is reasonable to set the lower integration limit to $-\infty$,
since our expressions for $A(E)$ and $\en(E,0)$ are valid for
$E\ll E_h$ and the contribution to the integral from
low energies falls off quickly with time as the most-bound particles are
eaten; hence any errors due to the forms of $A(E)$ or $\en(E)$ at
low energies are transient.
Setting the upper integration limit to $+\infty$ is also reasonable
since the product $A(E)\en(E,0)$ drops rapidly with $E$.
(Furthermore it is possible that the combined, luminous plus
dark matter density profiles in early-type galaxies are well
described as $\rho\sim r^{-2}$ even far outside of the nucleus.)
Making the substitution $y\equiv r_tt\exp({a+bE})$, we find
\begin{mathletters}
\begin{eqnarray}
\dot M(t) &=& {4\pi\sqrt{6} e\delta^3\over 9} e^{a/2} \overline{f_c}\left({r_t\over t}\right)^{1/2}\int_0^{\infty} y^{-1/2}e^{-y} dy \\
&\approx& \left(4.82,5.36\right)\overline{f_c}\left({r_t\over t}\right)^{1/2}
\end{eqnarray}
\end{mathletters}
where the paired numbers refer to $T=(0.5,0.25)$.
We note again the small
dependence of the feeding rate on the degree of triaxiality for
fixed $\overline f_c$.
The total mass accreted after time $t_{acc}$ is
\beq
\Delta M = \left(9.64,10.72\right)\overline{f_c}\left(r_tt_{acc}\right)^{1/2}.
\eeq
In physical units, these expressions become
\begin{mathletters}
\begin{eqnarray}
\dot M(t) &=& \left(1.48,1.54\right)\overline{f_c}{\sigma^3\over G} {\sigma\over c} \left({r_t\over r_s}\right)^{1/2}\left({t\over G\mh/\sigma^3}\right)^{-1/2} \\
&\approx&
\left(4.28, 4.46\right) \times 10^{-3}\msun {\rm yr}^{-1}\overline{f_c}\left({r_t\over r_s}\right)^{1/2}\left({\sigma\over 200\ {\rm km\ s}^{-1}}\right)^{5/2}\times \nonumber \\
& & \left({\mh\over 10^8\msun}\right)^{1/2}
 \left({t\over 10^{10}\ {\rm yr}}\right)^{-1/2}
\label{eq:emdot}
\end{eqnarray}
\end{mathletters}
and
\begin{mathletters}
\begin{eqnarray}
\Delta M &=& \left(2.97,3.09\right) \overline{f_c}\mh {\sigma\over c} \left({r_t\over r_s}\right)^{1/2}\left({t_{acc}\over G\mh/c^2}\right)^{1/2} \\ 
&\approx& 
\left(8.56, 8.92\right) \times 10^7 \msun \overline{f_c} \left({r_t\over r_s}\right)^{1/2}\left({\sigma\over 200\ {\rm km s}^{-1}}\right)^{5/2} \times 
\nonumber \\
& & \left({\mh\over 10^8\msun}\right)^{1/2}
\left({t_{acc}\over 10^{10}\ {\rm yr}}\right)^{1/2}
\label{eq:dm}
\end{eqnarray}
\end{mathletters}
The $r_t^{1/2}$ dependence reflects the more rapid destruction of
the cusp when $r_t$ is large.
Equation (\ref{eq:dm}) implies a captured mass of
order $\mh$ in a Hubble time.

\section{Tidal Disruption Rates}

The feeding rate in a steep-cusp nucleus 
is given by equation (\ref{eq:emdot}),
as a function of the capture radius $r_t$.
If $\mh\lap 10^8\msun$, solar type stars are tidally disrupted
at a radius
\beq
r_t \approx \left({\mh\over 10^8\msun}\right)^{-2/3} r_s
\label{eq:rt}
\eeq
\citep{Hills75}.
($10^8\msun$ is also roughly the mass of the largest
black holes that sit in steep-cusp nuclei.)
Combining equations (\ref{eq:emdot}) and (\ref{eq:rt}),
the tidal disruption rate in a steep-cusp nucleus is
\beq
\dot M_{collisionless}(t) \approx
4\times 10^{-3}\msun {\rm yr}^{-1}\overline{f_c}\ 
\sigma_{200}^{5/2}M_{\bullet,8}^{1/6} t_{10}^{-1/2}
\eeq
with $t_{10}$ the time since cusp formation in units of $10^{10}$ yr,
$\sigma_{200}$ the velocity dispersion in units of $200$ km s$^{-1}$ and
$M_{\bullet,8}$ the black hole mass in units of $10^8\msun$.
We refer to this as the ``collisionless'' tidal disruption rate
to highlight that the supply of stars to the black hole is not
being driven by gravitational scattering, as in the standard model
(Frank \& Rees 1976; Lightman \& Shapiro 1977).

Supermassive black holes in galaxies at the current epoch satisfy the
$\mh-\sigma$ relation,
\beq
\left({\mh\over 10^8\msun}\right) \approx 1.48 \left({\sigma\over 200\ {\rm km\ s}^{-1}}\right)^{4.65}
\eeq
(e.g. Merritt \& Ferrarese 2001b), which allows us to write
the tidal disruption rate in nearby galaxies
in terms of either $\mh$ or $\sigma$ alone:
\begin{mathletters}
\begin{eqnarray}
\dot M_{collisionless}(t) &\approx&
5\times 10^{-3}\msun {\rm yr}^{-1}\overline{f_c}\ 
\sigma_{200}^{3.28}\ t_{10}^{-1/2} \\
&\approx& 
4\times 10^{-3}\msun {\rm yr}^{-1}\overline{f_c}\ 
M_{\bullet,8}^{0.70}\ t_{10}^{-1/2}.
\end{eqnarray}
\end{mathletters}

We compare this to the rate at which collisional 
(encounter-driven) loss-cone
refilling supplies stars to the central black hole.
In a spherical, $\rho\propto r^{-2}$ nucleus,
\beq
\dot M_{collisional} =
2.8\times 10^{-4}\msun {\rm yr}^{-1}\ 
\sigma_{200}^{7/2}M_{\bullet,8}^{-1} 
\label{eq:WM}
\eeq
(Wang \& Merritt 2003);
this expression again assumes that the disrupted stars have the
mass and radius of the sun.
Equation (\ref{eq:WM}) can likewise be rewritten using the 
$\mh-\sigma$ relation:
\begin{mathletters}
\begin{eqnarray}
\dot M_{collisional}
&\approx& 
2\times 10^{-4}\msun {\rm yr}^{-1}
\sigma_{200}^{-1.15} \\
&\approx&
2\times 10^{-4}\msun {\rm yr}^{-1}M_{\bullet,8}^{-0.25}.
\label{eq:mcoll}
\end{eqnarray}
\end{mathletters}
Figure 5 plots $\dot M(\mh)$ in the collisionless (triaxial)
and collisional (spherical)
cases for a $\rho\sim r^{-2}$ nucleus.
The two rates scale in the opposite sense with $\mh$ and
collisional feeding would dominate in sufficiently small galaxies.
The rates are equal when
\beq
\mh \approx 5\times 10^6\msun \overline{f_c}^{-1}t_{10}^{0.5}.
\eeq
For $t_{10}^{1/2}/\overline{f_c}\approx 1$, 
this mass is close to that of the smallest black holes with reliably 
determined masses, in the Milky Way and M32.
Between this mass and $\mh\approx 10^8\msun$,
Figure 5 suggests that loss cone feeding driven by triaxiality
can easily dominate collisional loss cone feeding,
as long as the chaotic mass fraction $f_c$ is not too much smaller than one.
Flaring rates should peak in
the brightest galaxies with steep nuclear density profiles,
at values of a few times $10^{-3}\ \mathrm{yr}^{-1}$.
Since the disruption rate in the triaxial geometry
varies as $t^{-0.5}\propto (1+z)^{0.5}$,
flaring could be even more important at intermediate redshifts.

In a $\rho\propto r^{-1}$ nucleus, and again assuming $\mh\lap 10^8\msun$,
equation (\ref{eq:feed1}) gives a disruption rate for solar-type stars of
\beq
\dot M_{collisionless} \approx 2\times 10^{-5}\msun {\rm yr}^{-1} \overline{f_c} 
\left({r_h\over 100\ {\rm pc}}\right)^{-5/2}
\left({\mh\over 10^8\msun}\right)^{11/6}
\left({r_b\over 10 r_h}\right)^{5/8},
\eeq
approximately independent of time.
The corresponding collisional expression is 
\beq
\dot M_{collisional} \approx 1\times 10^{-6}\msun {\rm yr}^{-1} \overline{f_c} 
\left({r_h\over 100\ {\rm pc}}\right)^{-7/3}
\left({\mh\over 10^8\msun}\right)^{29/18}.
\label{eq:g1col}
\eeq
The functional dependence in equation (\ref{eq:g1col}) is taken from
Wang \& Merritt (2003), and the normalizing factor is based on
those authors' calculation of $\dot N$ for the galaxy NGC 3379
($\gamma\approx 1.1$).
For a $\gamma=1$ nucleus, the collisional and collisionless disruption
rates scale in almost the same way with $\mh$ and $r_h$, and
the collisionless rate exceeds the collisional rate as long as
$\overline{f_c}\gap 0.05$.

In summary: even small fractional populations of chaotic
orbits can produce tidal disruption rates that exceed those predicted
by the standard model of collisional loss cone repopulation
in a spherical nucleus (Syer \& Ulmer 1999; Magorrian \& Tremaine 1999;
Wang \& Merritt 2003).
In triaxial nuclei with $\rho\sim r^{-2}$ and $\mh\approx 10^8\msun$,
tidal flaring rates can plausibly reach values as high as several times
$10^{-3}$ yr$^{-1}$ for solar type stars.
Rates could be even higher if the cusps were recently formed.

\section{Black Hole Growth and the $\mh-\sigma$ Relation}

The fate of gas liberated by the tidal disruption of a star at
$r_t>r_s$ is uncertain.
But for $\mh\gap 10^8\msun$, stars are directly captured by the black hole
without being disrupted.
Using equation (\ref{eq:dm}), setting $r_t=r_s$ and 
requiring $\Delta M = \mh$, we find the following relation between
$\mh$ and $\sigma$ in a steep triaxial cusp:
\beq
{\mh\over 10^8\msun} \approx 0.8 \overline{f_c} \left({t_{acc}\over 10^{10}\ {\rm yr}}\right) \left({\sigma\over 200\ {\rm km\ s}^{-1}}\right)^5.
\eeq
This is remarkably similar to the $\mh-\sigma$ relation: 
the exponent on $\sigma$ is consistent with measured values,
$4.5\pm 0.5$ \citep{MF01a}
and even the normalization is of the right order if $\overline{f_c}\approx 1$.

The $\dot M\propto\sigma^5$ dependence can be understood in several
ways.
Writing $\dot\en(E,0)=r_sA(E)\en_c(E,0)$ 
and integrating over energies from $E_h$ to infinity,
we find
\beq
\dot M(E\ge E_h) \approx 2.5 f_c {\sigma^3\over G}\left({\sigma\over c}\right)^2.
\label{eq:emdot2}
\eeq
Alternatively, 
the total rate at which stars pass inward at radius $r$ is
$\sim 4\pi r^2\rho(r)\sigma=2\sigma^3/G$,
and a fraction $F=T_D A f_c r_s$
of stars at energy $E$ pass within $r_s$ 
each crossing time.
Equation (\ref{eq:aofe}) gives
\beq
A(r) \propto {\sigma^5\over G^2\mh^2}\left({r\over r_h}\right)^{-2}
\eeq
and $T_D\propto (r_h/\sigma)(r/r_h)$, so that
\beq
\dot M(r) \propto \overline{f_c} {\sigma^3\over G}\left({\sigma\over c}\right)^2{r_h\over r}.
\label{eq:sig5}
\eeq
These relations show that the feeding rate due to stars originating
outside of the SBH's sphere of influence is of order
\beq
{\sigma^5\over G c^2} \approx 10^{-3}\msun \mathrm{yr}^{-1} \left({\sigma\over 200\ \mathrm{km}\ \mathrm{s}^{-1}}\right)^5,
\label{eq:approx}
\eeq
independent of $\mh$ -- large enough to contribute substantially
to the masses of SBHs if accretion continues for $10^9$ yr
or more.

For $\mh\lap 10^8\msun$, the tidal disruption radius exceeds
$r_s$ and scales as $\mh^{1/3}$ (equation \ref{eq:rt}).
If the growth of the black hole is determined by the rate
of tidal disruptions,
$\Delta M$ scales more weakly with $\sigma$ in this regime, 
roughly as $\sigma^3$.
This would imply a flattening of the $\mh-\sigma$ relation at masses
below $\sim 10^8\msun$.
However it is not clear what the fate of tidally-liberated gas would
be; some would fall into the black hole but some would escape
\citep{Frank79}.
For masses below $\sim 10^6\msun$, Figure 5 suggests that
collisional loss cone repopulation would dominate over
collisionless feeding.

Equation (\ref{eq:approx}) is similar to 
equation (8) of Zhao, Haehnelt \& Rees (2002), who considered 
the consequences for black hole growth of a continuously
resupplied loss cone in the spherical geometry.
The agreement is reasonable since the feeding rates in the
triaxial models are comparable to spherical, full-loss-cone rates 
for $E\gap E_h$ (Figure 3).
The triaxial models are different in one respect.
In the spherical geometry, the contribution per orbital period 
to $\Delta M$ from stars at radius $r$ scales as $r^{-2}$.
In the triaxial models, every centrophilic orbit is capable
of visiting the center and the probability that a single star
will do so in one orbital period varies as $\sim r^{-1}$.
Thus while the feeding rates from stars at $r\approx r_h$ are comparable
in the two models, large-radius orbits contribute relatively more
in the triaxial models.
One consequence is that the growth of the black hole is not so strongly
dependent on the stellar distribution near $r_h$.

Accretion of stars,
or of gas liberated from stars,
have long been discussed as mechanisms for growing SBHs
in galactic nuclei 
(Hills 1975; Young, Shields \& Wheeler 1977; Frank 1978; 
McMillan, Lightman \& Cohn 1981).
Hills pointed out already in 1975 that feeding at the full-loss-cone
rate could plausibly grow a $10^8\msun$ black hole in a Hubble time.
Subsequent authors (e.g. Young, Shields \& Wheeler) tended
criticize this claim on the grounds that gravitational encounters
occur too infrequently to maintain a fully-populated
loss cone.
Our result -- that sustained feeding rates in the triaxial geometry 
can be comparable to full-loss-cone rates in the spherical geometry --
should help to revive Hills' model.
However, chaotic loss cones still fail to reproduce in a natural
way the observed time dependence of quasar luminosities.
The $t^{-1/2}$ time dependence found here for stellar feeding
in a $\rho\sim r^{-2}$ density cusp (equation \ref{eq:emdot}) --
which results entirely from the depopulation of chaotic orbits --
is much more gradual than the observed,
$\sim (1+z)^3$ falloff of quasar luminosities
(e.g. Boyle, Shanks \& Peterson 1988), and would require 
that much of the growth took place long after the quasar epoch.
Slow, optically faint growth of SBHs has been proposed
(Richstone et al. 1998),
but only before it was discovered that SBH masses in nearby galaxies
had been substantially overestimated \citep{MF01b}.
SBH masses are now
consistent with accretion rates inferred from quasar statistics
\citep{MF01c}
leaving less room for growth at late times.

On the other hand, the time dependence derived here assumes
that the steep density cusp forms only once and that it
is slowly depleted by accretion of centrophilic orbits.
Cusps might form episodically via star formation
during mergers leading to repeated bursts of stellar fueling
and larger average rates of accretion \citep{MM03b}.
Other considerations could
boost the mean capture rate as well.
Growth of the SBH via accretion steepens the
stellar density profile, to $\rho\sim r^{-2.5}$ near the SBH
\citep{Merritt03},
an effect that was ignored here.
Depopulation of the chaotic orbits would be counteracted to
some extent by scattering of stars onto these orbits.
Dark matter could also be accreted (Zhao, Haehnelt \& Rees 2002;
Read \& Gilmore 2003).
Once the supply of star-forming gas was depleted
\citep{KH00}, subsequent mergers would turn off the stellar
feeding through the formation of a binary SBH which ejects stars from the
nucleus and lowers its density \citep{MM01}.
This might be a natural way to explain the rapid observed falloff in quasar
luminosities.
In summary, 
while capture of stars is probably not the only way in which black 
holes grow,
there would seem to be no reason to rule it out 
as an important contributor.

\section{Transient Changes in the Potential}

Deviations from axial symmetry are likely to occur
at least temporarily following a merger or accretion event,
yielding chaotic orbits near the black hole and
resulting in enhanced feeding rates \citep{NS83}.
Assuming a steep cusp (perhaps formed dissipatively during the merger), 
equation (\ref{eq:dm}) and the $\mh-\sigma$
relation imply an accreted mass of
\beq
\Delta M \approx 2\times 10^7\msun \overline{f_c}\left({r_t\over r_s}\right)^{0.5} \left({\sigma\over 200\ {\rm km\ s}^{-1}}\right)^5\left({\Delta t\over 10^8\ {\rm yr}}\right)^{0.5}
\eeq
in time $\Delta t$, or a mean feeding rate over $\Delta t$ of
\beq
{\Delta M\over\Delta t} \approx 0.2 \msun\ {\rm yr}^{-1} \overline{f_c}\left({r_t\over r_s}\right)^{0.5} \left({\sigma\over 200\ {\rm km\ s}^{-1}}\right)^5\left({\Delta t\over 10^8\ {\rm yr}}\right)^{-0.5}.
\label{eq:fuel}
\eeq
This is comparable to feeding rates inferred in AGNs for
$\sigma\lap 250$ km s$^{-1}$ and in quasars for
$\sigma\gap 350$ km s$^{-1}$.

\section{Decay of a Black-Hole Binary}

Following the merger of two galaxies each containing a SBH,
a binary forms with semi-major axis $a=a_{hard}$, where
\begin{mathletters}
\begin{eqnarray}
a_{hard} &\approx& {G\mu\over 4\sigma^2} \approx 2.7\ {\rm pc} (1+p)^{-1} \left({m_2\over 10^7\msun}\right)\left({\sigma\over 200\ {\rm km\ s}^{-1}}\right)^{-2} \\
&=& {p\over 4(1+p)^2} r_h;
\end{eqnarray}
\end{mathletters}
$m_2$ is the mass of the smaller black hole, $p\equiv m_2/m_1$ and
$\mu\equiv m_1m_2/\mh$ \citep{Merritt03}.
In a spherical or axisymmetric galaxy, the binary quickly ejects stars with
pericenters below $\sim a_{hard}$, after which changes in the binary separation
take place on the much longer time scale associated
with collisional refilling of the binary's loss cone \cite{Yu02}.
In all but the densest nuclei, decay stalls at a separation
$a \gap 0.1 a_{hard}$, too large for the efficient emission
of gravitational waves \citep{Merritt03}.
In a triaxial galaxy, decay can continue due to the much larger
supply of centrophilic stars.
The decay rate is determined by the supply of stars into
a region of radius $\sim a$ around the center of mass of the 
binary \citep{Quinlan96}.
In a steep triaxial cusp, the feeding rate from stars of
energy $E>\Phi(a_{hard})$ into a region of
radius $\sim a_{hard}$ is 
given by an equation similar to (\ref{eq:emdot2}):
\begin{mathletters}
\begin{eqnarray}
\dot M(E>E_{hard})&\approx& a_{hard}\int_{E_{hard}}^{\infty} A(E)M_c(E) dE \\
&\approx& 1.3\overline{f_c} {\sigma^3\over G} \\
&\approx& 
 2500\msun\ {\rm yr}^{-1}\overline{f_c} \left({\sigma\over 200\ {\rm km\ s}^3}\right)^3,
\end{eqnarray}
\end{mathletters}
independent of $\mh$ and $p$.
If this rate were maintained, the binary would interact with
its own mass in stars in a time of only $\sim 10^5$ yr.
In fact, the feeding rate will decline with time as the 
centrophilic orbits are depleted.
Equating the energy carried away by stars with the change in the binary's
binding energy gives
\beq
{3\over 2}{G\mu\over a} dM\approx {Gm_1m_2\over 2} d\left({1\over a}\right)
\eeq
\citep{Merritt03}.
The coupled equations describing the change in the binary separation and 
the evolution of the stellar distribution are then
\begin{mathletters}
\begin{eqnarray}
{d(1/a)\over dt} &=& {3\over \mh} {r_t\over a} \int A(E)\en_c(E,t) dE, 
\label{eq:evolve1} \\
{d\en_c\over dt} &=& -r_t A(E) \en_c(E,t)
\label{eq:evolve2}
\end{eqnarray}
\end{mathletters}
where $r_t$ is the pericenter distance below which stars can
efficiently exchange energy with the binary.
Note that we are ignoring any contribution to shrinkage of the
binary from low-angular-momentum regular orbits, or from
collisional repopulation of the loss cone.

We set $r_t$ to a (small) multiple of $a$, 
$r_t(t)={\cal R} a(t)$, ${\cal R}\approx 1$, reflecting the fact
that only stars which come to within roughly a binary
separation of the black holes will interact strongly enough
with them to be ejected.
As lower integration limit on the integral in equation (\ref{eq:evolve1}),
we take $E_{hard}$; this choice reflects the fact that the
stellar distribution immediately following the formation of 
a hard binary is similar to that in the pre-merger galaxies outside
of $r\approx a_{hard}$ \citep{MM01}.
The evolution equations become
\begin{mathletters}
\begin{eqnarray}
{d(1/a)\over dt} &=& {3\over \mh} {\cal R} \int_{E_{hard}}^\infty
A(E)\en_c(E,t) dE, 
\label{eq:evolvea} \\
{d\en_c\over dt} &=& -{\cal R} a(t) A(E) \en_c(E,t).
\label{eq:evolveb}
\end{eqnarray}
\end{mathletters}
As initial conditions we take $a(0)=a_{hard}$ and 
$\en_c(E,0)=\overline{f_c}\en(E)$ with $\en(E)$ given by 
equation (\ref{eq:nofe2}).

Figure 6 shows solutions for various choices of $\overline{f_c}$
and for $p\equiv m_2/m_1=(1,0.1)$.
At late times, the binary separation varies approximately as
\beq
{a_{hard}\over a} \approx \overline{f_c}^2 \times t
\eeq
in model units, for both values of $p$.
In physical units,
\beq
{a_{hard}\over a} \approx 3\times 10^4 \overline{f_c}^2 \left({\sigma\over 200\ \mathrm{km}\ \mathrm{s}^{-1}}\right)^3 \left({\mh\over 10^8\msun}\right)^{-1} \left({t\over 10^{10} \mathrm{yr}}\right). 
\label{eq:aoft}
\eeq
A $1/a\propto t$ dependence is expected in ``full loss cone''
situations like this one (Milosavljevic \& Merritt 2003b).
The nonlinear dependence of $a(t)$ on $\overline{f_c}$ reflects the
fact that, when $\overline{f_c}$ is small, the decay rate is also small
and the binary tends to stall at a large separation,
causing it to interact with and deplete the supply of
stars, further slowing the decay etc.

Coalescence due to gravitational wave emission in a time 
$t_{gr}$ occurs when $a=a_{gr}$, where
\begin{equation}
{a_{hard}\over a_{gr}} \approx 75 {p^{3/4}\over\left(1+p\right)^{3/2}} 
\left({\sigma\over 200\ {\rm km\ s}^{-1}}\right)^{-7/8} 
\left({t_{gr}\over 10^9{\rm yr}}\right)^{-1/4}
\label{eq:a_gr}
\end{equation}
for a circular-orbit binary \citep{Merritt03}.
For $p= 1 (0.1)$, equation (\ref{eq:a_gr}) implies that
decay by a factor of $\sim 10^2 (10^1)$ is required
in order for gravity-wave coalescence to occur in $10^9$ yr.
Combining equations (\ref{eq:aoft}) and (\ref{eq:a_gr}),
this can be achieved in $10^{10}$ yr
for chaotic mass fractions:
\beq
\overline{f_c} \approx 0.05 \left({\sigma\over 200\ \mathrm{km}\ \mathrm{s}^{-1}}\right)^{9/16}
\eeq
with a weak dependence on $p$.
Thus, placing just a few percent of a galaxy's mass on chaotic orbits
is sufficient to overcome the ``final parsec problem'' 
\citep{MM03a} and induce coalesence,
{\it if} (as assumed in Figure 6) the chaotic orbits are present
initially at energies much greater than $E_h$.
Since the stars supplied to the binary come from well outside
of $r_h$ in this case, the effect on the stellar density profile 
would be negligible
even after the binary had ejected several times its own mass in stars.

\section{Summary}

In a $\rho\sim r^{-2}$ triaxial nucleus containing chaotic 
(centrophilic) orbits and a central
black hole, stars are supplied to the black hole's sphere
of influence at a rate $\sim f_c\sigma^3/G$, and to
the black hole itself at a rate $\sim f_c\sigma^5/Gc^2$,
where $f_c$ is the fraction of the mass on centrophilic
orbits and $\sigma$ is the stellar velocity dispersion.
If $f_c$ is of order unity, as in recently published self-consistent
models (Papers II, III), the mass accumulated in $10^{10}$ yr
is of the same order as observed black hole masses.
Feeding rates in such an environment fall off as $\sim t^{-1/2}$
as the chaotic orbits are depleted.
Tidal disruption rates at the current epoch due to stars on chaotic
orbits are as much as 1-2 orders of magnitude greater than in
standard, spherical or axisymmetric models in which loss-cone refilling
occurs via gravitational scattering onto eccentric orbits.
Transient changes in the shape of a nucleus during mergers
or accretion events could lead to episodic stellar feeding at rates
of $\sim 0.2-2\msun$ yr$^{-1}$, comparable to the rates inferred
in active galactic nuclei and quasars.
Decay of a black-hole binary at the center of a triaxial nucleus
could be greatly enhanced compared with spherical or axisymmetric
nuclei, even if only a few percent of the mass is on chaotic
orbits, thus solving the ``final parsec problem.''

\bigskip

This work was supported by NSF grants
AST 00-71099 and AST 02-0631,
and by NASA grants NAG5-6037 and NAG5-9046. 
M.Y. Poon is grateful to the Croucher Foundation for a postdoctoral 
fellowship.

\myputfigure{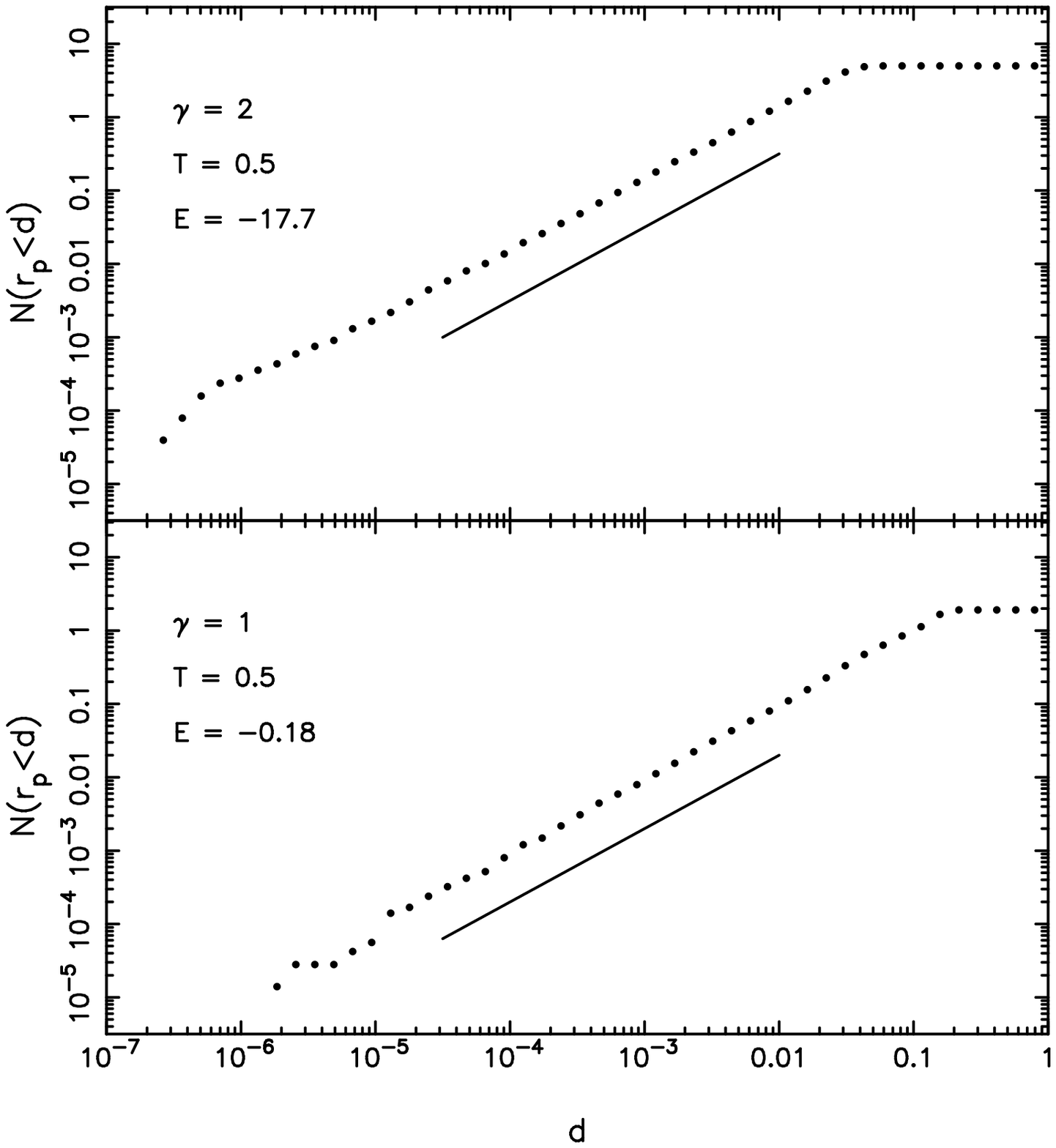}{6.}{0.75}{-90}{-50}
\figcaption{\label{fig:1}
Number of central encounters per unit time with pericenter distance
less than $d$, for two chaotic orbits, each integrated for $10^5$
dynamical times in their respective potentials. The solid lines
have unit slope.}
\vspace{\baselineskip}

\myputfigure{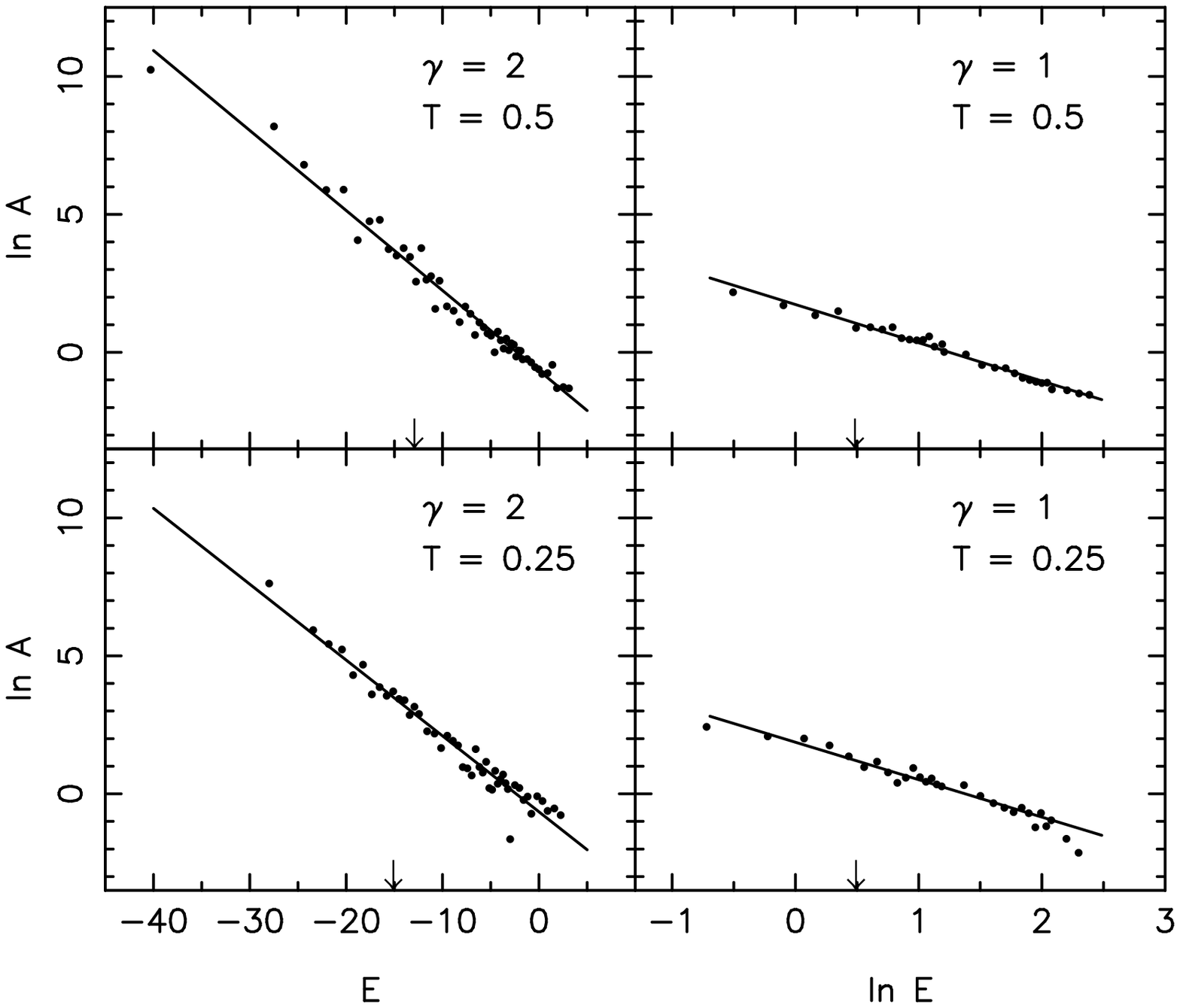}{6.}{0.75}{-100}{-50}
\figcaption{\label{fig:2}
The function $A(E)$ that describes the cumulative rate of pericenter
passages in the four triaxial models.
Points are from integrations of chaotic orbits; lines show the fits
described in the text.
Arrows indicate $E_h$.
}
\vspace{\baselineskip}

\myputfigure{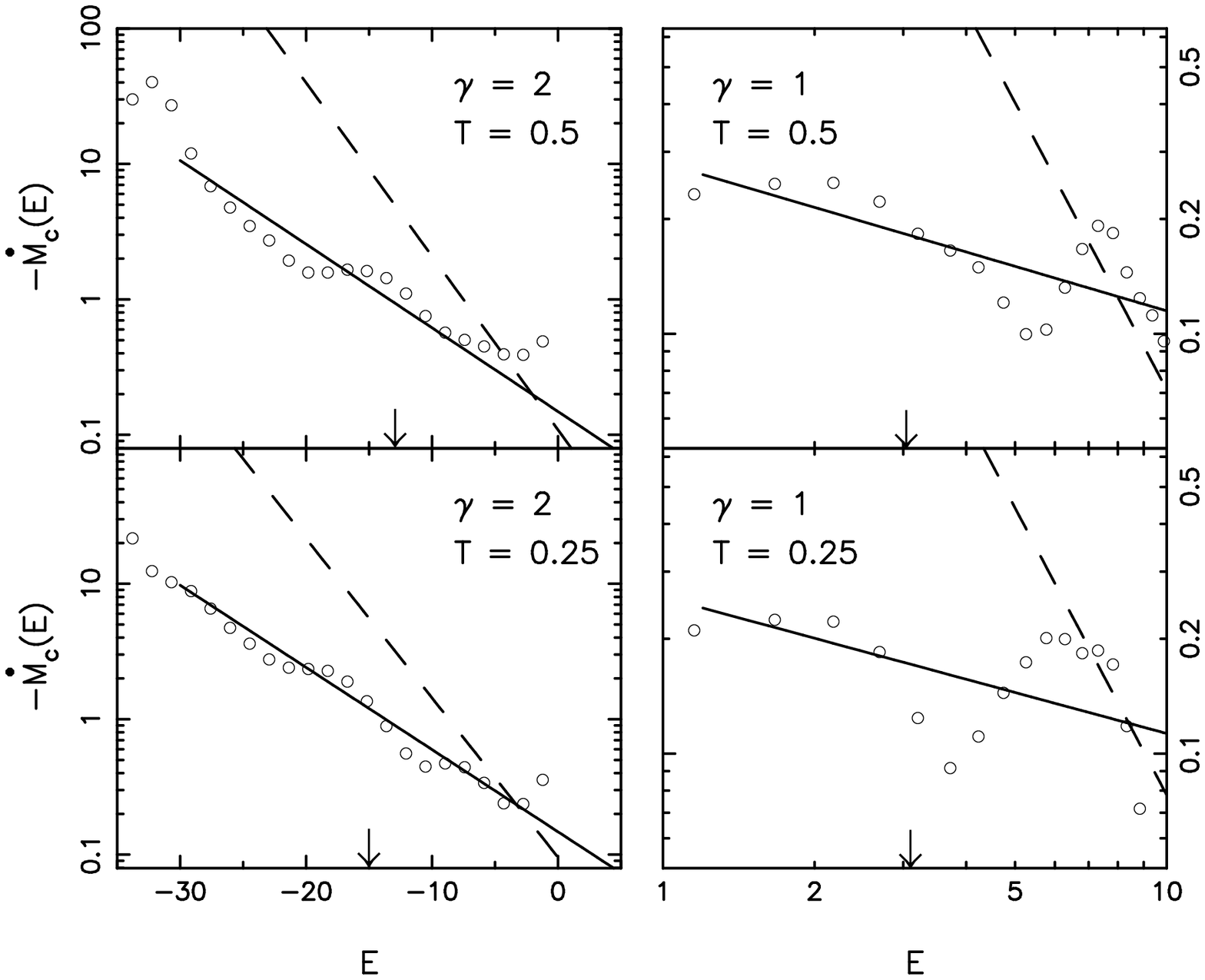}{6.}{0.75}{-100}{-50}
\figcaption{\label{fig:3}
Energy-dependent capture rate $\dot\en_c(E)$ due to chaotic orbits
in the four triaxial models at $t=0$.
The capture radius $r_t$ has been set to unity; capture
rates scale linearly with $r_t$ (cf. eq. \ref{eq:capture}).
Open circles: $\dot\en_c$ computed using the actual chaotic orbit
populations in the Schwarzschild models.
Solid lines: $\dot\en_c$ computed using the analytic approximations
to $\en_c$ described in the text.
Dashed lines: Capture rates predicted by a spherical, full-loss-cone
model.
Arrows indicate $E_h$.
}
\vspace{\baselineskip}

\myputfigure{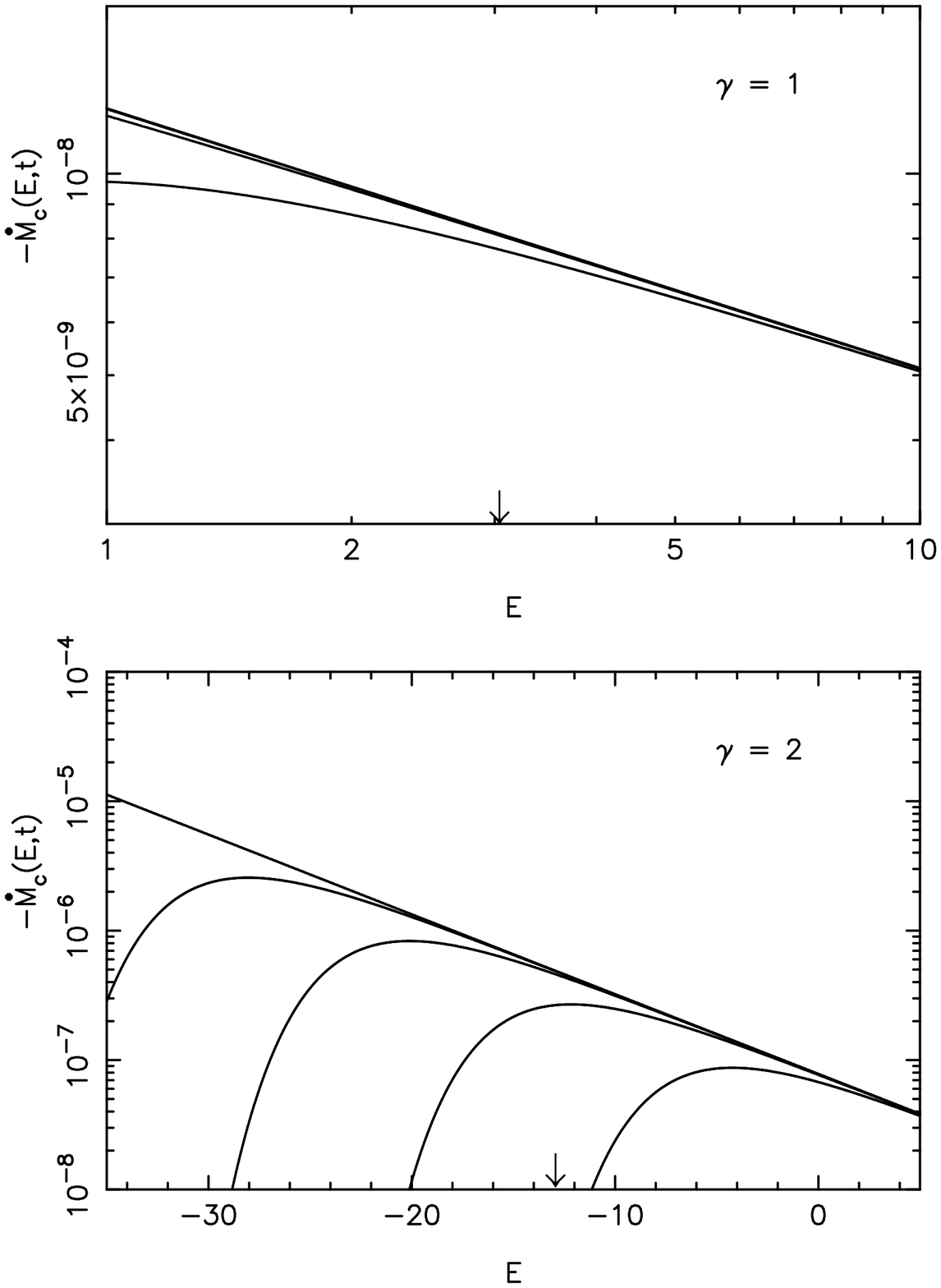}{6.}{0.75}{-20}{-50}
\figcaption{\label{fig:4}
$\dot\en_c(E,t)$ at four times, $t=(0,10^3,10^4,10^5,10^6)$ in
model units, for $\gamma=1$ and $2$, and $T=0.5$.
The capture radius $r_t$ has been set to the Schwarzschild radius
$r_s$ using the scalings discussed in the text.
Arrows indicate $E_h$.
}
\vspace{\baselineskip}

\myputfigure{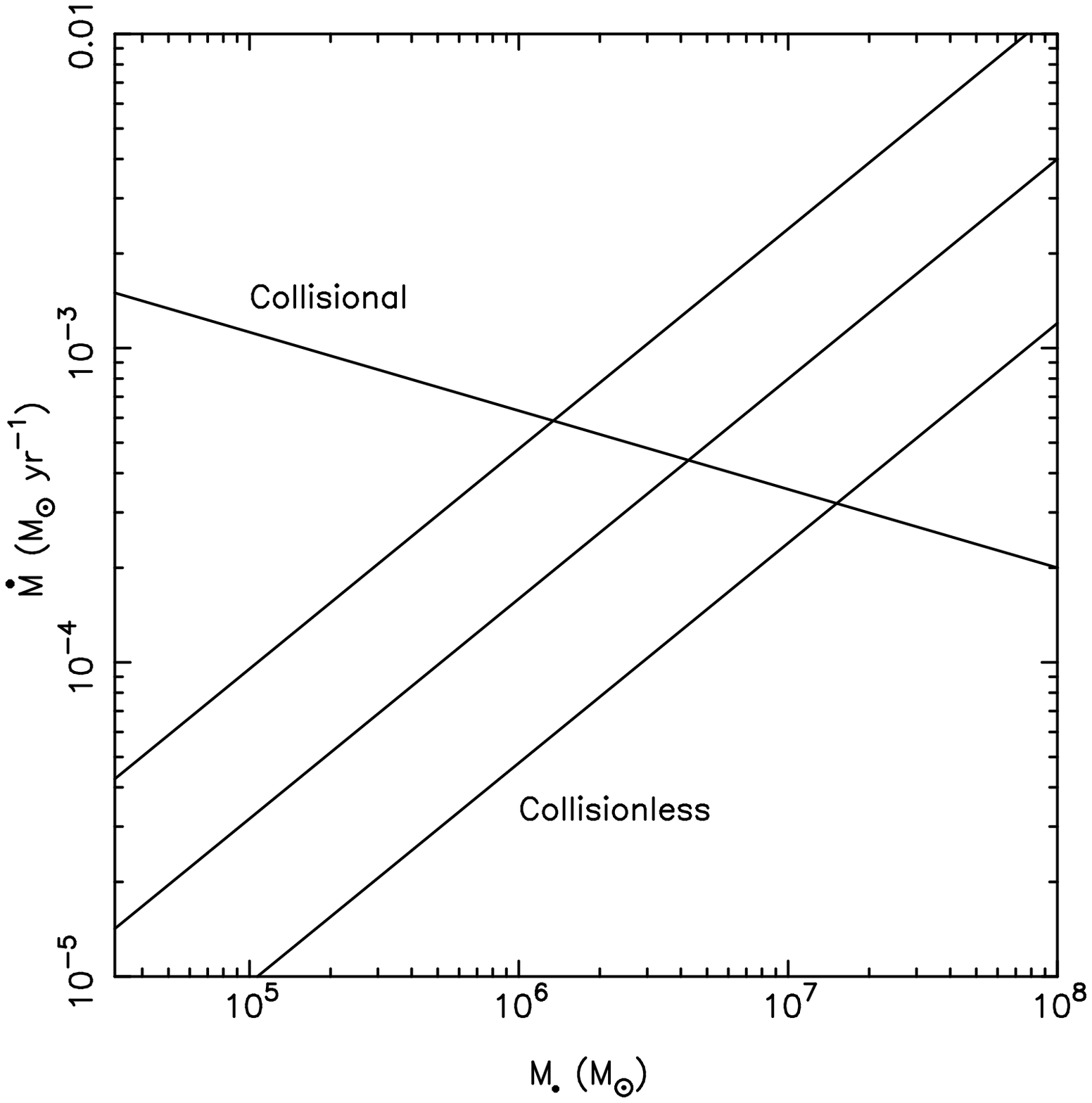}{6.}{0.75}{-20}{-50}
\figcaption{\label{fig:6}
Tidal disruption rates for solar type stars in collisional (spherical)
and collisionless (triaxial) nuclei with $\rho\propto r^{-2}$.
Line labelled ``collisional'' is equation (37).
``Collisionless'' lines are equation (35)
with $\overline{f_c}t_{10}^{-1/2}=(0.3,1,3)$.
}
\vspace{\baselineskip}

\myputfigure{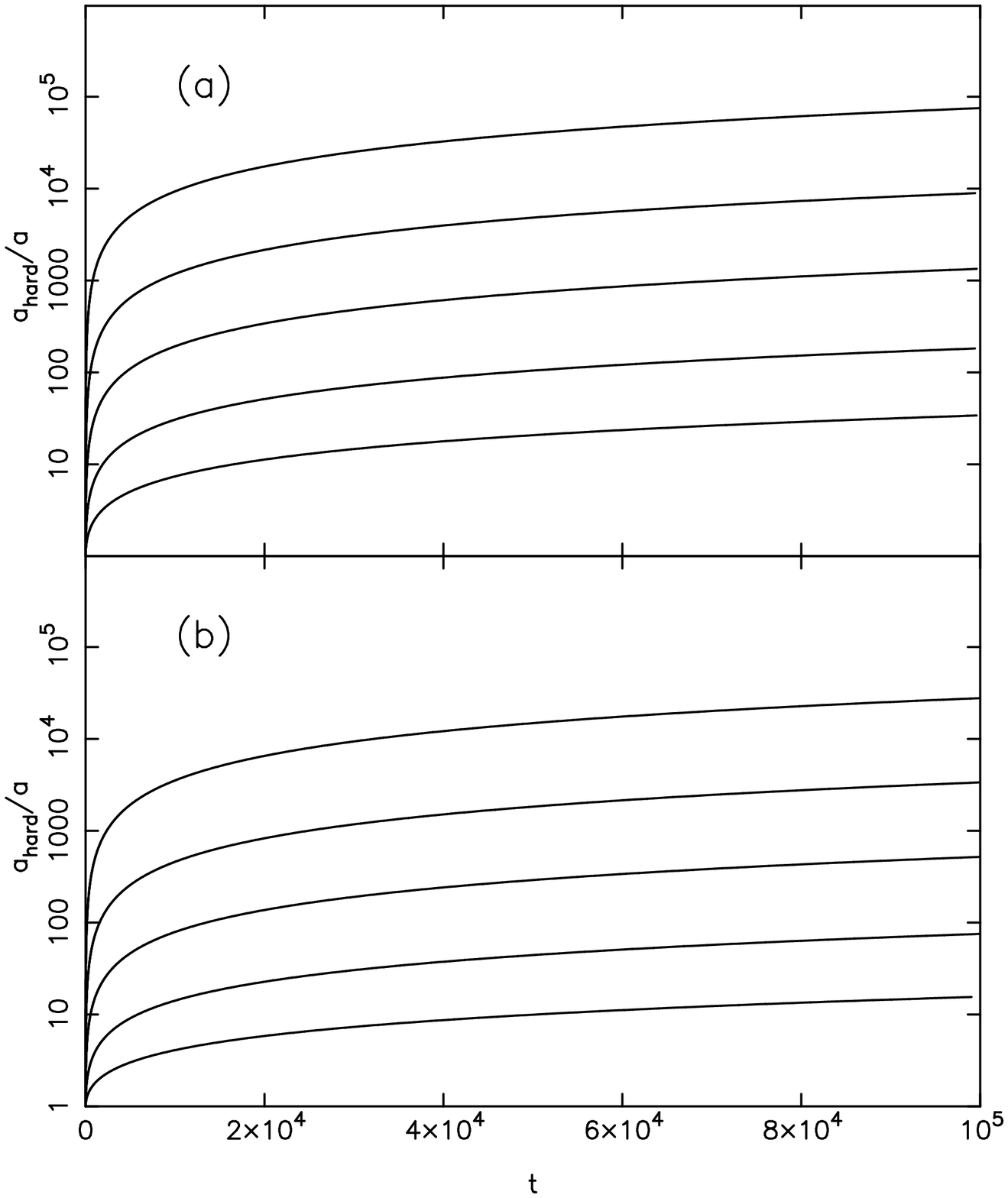}{6.}{0.75}{-20}{-50}
\figcaption{\label{fig:5}
Decay of a massive binary black hole at the center of
a steep-cusp ($\rho\sim r^{-2}$) nucleus due to ejection of
stars on chaotic orbits.
(a) $m_2/m_1=1$; (b) $m_2/m_1=0.1$.
The different curves are for chaotic mass fractions
of $\overline{f_c}=1$ (top), $0.3,0.1,0.03$ and $0.01$ (bottom).
Time is measured in model units (equation \ref{eq:defT2}); 
$t=10^5$ corresponds roughly to $10^{10}$ yr.
}
\vspace{\baselineskip}

\end{document}